# Machine learning-based determination of magnetic parameters from magnetic images with different imaging scales

Akito Watanabe(渡辺章斗)[1], Yoshinobu Nakatani(仲谷栄伸)[1], Hiroyuki Awano(粟野博之)[1], and Kenji Tanabe(田辺賢士)[1, a)]

[1]Toyota Technological Institute, Nagoya, 468-8511, Japan

[2]University of Electro-Communications, Tokyo, 182-8585, Japan

[a)]Author to whom correspondence should be addressed: tanabe@toyota-ti.ac.jp

The determination of material parameters is significantly important in material science, which is often a challenging task. Recently, advancements have shown that magnetic parameters, such as the Dzyaloshinskii–Moriya interaction (DMI), can be estimated from a magnetic domain image using machine learning (ML). This development suggests a potential shift in how magnetic parameters are determined, moving away from traditional measurement techniques to more innovative methods involving image-based inputs processed by ML. In previous studies, the test images used for estimation always matched the training images in size. However, since image size is contingent on the microscopy technique used, the ability to accurately estimate parameters from images of varying sizes is essential. Here, we investigated the feasibility of estimating the DMI constant and saturation magnetization from magnetic domain images of different sizes using ML. We successfully demonstrated that it is possible to estimate these parameters even when the imaging sizes differ between training and test datasets. Additionally, our comparison of the estimation accuracy for the DMI constant and saturation magnetization revealed that the tolerance for differences in image size varies depending on the specific parameter being estimated. These findings could have a significant impact on the future methods of determining magnetic parameters.

In the field of spintronics, research typically progresses through the fabrication of magnetic films, evaluation of their magnetic properties, and device processing. Evaluating magnetic properties involves measuring various magnetic parameters, such as saturation magnetization and magnetic anisotropy. However, certain measurements, like those for exchange stiffness constant[1-6] and Dzyaloshinskii-Moriya interaction (DMI) constant[7-16], can be particularly challenging or time-consuming, although the DMI constant is a particularly important parameter for research into next-generation memory devices[17-22]. To enhance research efficiency, there is a growing need for simple and rapid measurement methods[23].

Recently, the use of machine learning to determine magnetic parameters from magnetic domain images has emerged as a promising approach[24-30]. Magnetic domain images are formed in a way that minimizes the total magnetic energy[31], suggesting that even in complex or random magnetic domain structures, there may be embedded information about the underlying magnetic parameters. By training artificial intelligence (AI) models on a large dataset of magnetic domain images, it becomes possible to estimate these parameters simply by showing magnetic domain images to the trained AI as shown in Fig. 1(a).

Machine learning methods, however, typically require a large amount of training data, which in this context means a substantial number of magnetic domain images. Experimentally generating such images is often impractical, so researchers rely on numerical methods like micromagnetic simulations to produce the training data. While smaller images are easier and less computationally expensive to generate in simulations (the left side of Fig. 1(a)), actual test data—magnetic domain images obtained from various microscopes—may vary in size depending on the imaging equipment used. Generally, microscopes with lower resolution capture larger images (the right side of Fig. 1(a)), leading to a potential mismatch in image sizes between the training data and test data.

Previous studies typically ensured that the training data and test data were same in size. However, it is not always feasible to produce images of the same size, and ideally, the AI should be able to estimate magnetic parameters even when the training and test images differ in size. In this study, we explored whether it is possible to accurately estimate magnetic parameters, such as the DMI constant and saturation magnetization, despite differences in the imaging scale between training and test datasets. Our findings revealed that under certain conditions, it is indeed possible to estimate these parameters even when the image sizes are not same. This suggests that when AI estimates magnetic parameters, it focuses not on features that vary with curvature or size changes but rather on structural patterns that remain consistent regardless of image scale. Additionally, our comparison between the estimation of the DMI constant and saturation magnetization revealed that the tolerable difference in image size varies depending on the parameter being estimated. Parameters that more significantly influence features sensitive to size changes, such as curvature and magnetic domain size, tend to have a lower tolerance for differences in image size. These insights could significantly enhance the flexibility and applicability of machine learning methods in the evaluation of magnetic properties, making it easier to integrate data from diverse sources and imaging conditions.

All magnetic domain images were made by micromagnetic simulations of using a GPU-based program developed previously[40-41]. The details are written in the previous articles[26]. The materials parameters used in the calculations are summarized in Table 1. All the magnetic films are perpendicularly magnetized. The dimension of the cell is $4 \times 4 \times 4 \, nm^3$. The thickness of the ferromagnetic layer is 4 nm and the single cell layer is used. DMI is introduced between nearest neighbor cells and can be considered as bulk DMI. The number of the cell is $512 \times 512$ (the image size is $2.048 \times 2.048 \, \mu m^2$). The periodic boundary condition is employed to avoid effects from the edges. 10,000 (5,000) simulations were carried out to produce the results presented in Figs. 2-3(4-5). Of the images generated, 80% are used for training and the remaining 20% are used for test.

The image size of $2.0 \times 2.0 \, \mu m^2$ was changed to different image sizes such as $0.5 \times 0.5, \; 1.0 \times 1.0, \;$ and $1.5 \times 1.5 \, \mu m^2$ by cutting. At this time, the amount of data also increases due to cutting. Each image is converted to a $128 \times 128$ pixels image (one-pixel size depends on the image size) using an averaging filter. For data augmentation, using rotation (90, 180, 270 deg.) increases the number of images by four times as shown in Fig. 1(b).

The CNN system used in this paper contains seven layers. The first six are convolution layers, including ReLU, Batch Normalization and Max Pooling layers, and the remaining one is fully connected. The filter size of the convolution layers is $3 \times 3$ and the number of the filters is 16, 32, 64, 64, 128, and 128. The filter size of the max pooling layer is $2 \times 2$. A batch size is 128 and numbers of epochs are from 200 to 400, depending on the convergence of the loss function. The network was trained using a commercial deep learning tool, Sony Neural Network Console (https://dl.sony.com/app/).

Firstly, we conducted an experiment to estimate the DMI constant using images with one imaging scale as training data and images with a different imaging scale as test data. The training data consist of images with three different scales: 0.5, 1.0, and 1.5 μm, while the test data was an image with a 2.0 μm scale (Fig. 2(a)). Figure 2(b) shows the estimated results. For the 1.5 μm scale, a proportional relationship was observed between the set DMI value and the estimated DMI value, indicating that the imaging scale of the training data and test data does not necessarily need to be identical to estimate the DMI constant. There are larger errors at the edges of the data, which we will discuss later. For the 0.5 and 1.0 μm scales, however, the estimation is not successful. Note that although we conducted experiments with different amounts of data from 40,000 to 640,000, no significant differences were observed. Whether estimation is possible does not depend on the amount of data.

There are two possible explanations for the result on the unsuccessful estimation. The first is that the images with 0.5 and 1.0 μm may not contain discernible DMI features due to the small scale. The second possibility is that the difference in imaging scale between the training data and test data was too large. To investigate this further, we used images at 0.5 and 1.0 μm for both the training and test data as shown in Fig. 2(c-d). In both cases, linear relationships were observed, indicating successful estimation. This result suggests that the first possibility, which posits that the DMI features were absent from the magnetic domain images at smaller scales, is not the correct explanation. Instead, it appears that the discrepancy in scale between the training and test data is the more likely cause of the unsuccessful estimations in the initial experiments.

Secondly, we conducted an experiment to estimate the DMI constant using images with multiple image scales for both the training data and the test data. We used images with scales of 0.5, 1.0, 1.5, and 2.0 μm for both the training and test data (Test A in Fig. 3(a)). The results indicate that the estimation was successful for all image sizes as shown in Fig. 3(b). This finding demonstrates that it is not necessary for the image sizes of all training and test data to match exactly. Moreover, we conducted an experiment where the test data consisted of images with scales (0.75, 1.25, and 1.75 μm) that were not included in the training data (Test B in Fig. 3(a)). As shown in Fig. 3(c), a proportional relationship was also observed in this case. This result reveals that high accuracy in estimation can be achieved even when the training data does not contain images of the same size as the test data. In images with different imaging scales, the curvature and other length scales of the domain wall shape appear significantly different. Nevertheless, the successful estimation across varying scales suggests that the AI estimates DMI based on similar domain wall shapes that remain consistent despite resizing like fractal, rather than relying solely on curvature or specific scale-dependent features.

Thirdly, we conducted an experiment to estimate not only the DMI value but also the saturation magnetization. Similar to the experiment in Fig. 3, magnetic domain images with multiple image sizes were used for training, and two types of data sets were tested. Figure 4(b) shows the results when using the same image sizes as the training data (0.5, 1.0, 1.5, and 2.0 µm) for the test data. As with the DMI estimation experiment, the estimation was successful overall. For saturation magnetization, however, the estimation accuracy decreased in regions with low saturation magnetization.

This decrease in accuracy may be due to the reduction of the influence of magnetostatic energy. As illustrated in Fig. 4(c), when saturation magnetization is low, the influence of the magnetostatic energy decreases, leading to a reduction in the area occupied by domain walls in order to minimize the exchange energy and magnetic anisotropic energy (commonly referred to as domain wall energy). As a result, the magnetic domain size increases. Since the shape of the domain wall carries critical information about the magnetic parameters such as saturation magnetization, the increase in magnetic domain size may lead to a loss of detailed information on the magnetic parameters, thereby reducing the accuracy of the estimation. Additionally, the reduced influence of magnetostatic energy on magnetic domain states may lead to the loss of subtle information embedded in these states. Conversely, in regions with high saturation magnetization, the estimation accuracy was better than that observed in the DMI estimation. In the DMI estimation, accuracy typically declined at the extremities of the DMI value range, but this did not occur to the same extent in saturation magnetization estimation. This suggests that factors other than the quantity of training data at the edge points might be contributing to the accuracy variations in DMI estimation.

Furthermore, we conducted estimation experiments using test data with image sizes of 0.75, 1.25, and 1.75 µm, which were different from those in the training data. In this case, significant errors were observed in Fig. 4(d), unlike in the DMI estimation. The most notable errors occurred at the 0.75 µm scale, where the results were not well estimated. This indicates that while the model could generalize well for DMI across different scales, it struggled with saturation magnetization, particularly when the test image sizes did not match those in the training set.

To clarify why the estimation of saturation magnetization from the magnetic domain image with an image size of 0.75 μm was unsuccessful, we conducted experiments under varying conditions. First, to determine whether the image size of 0.75 μm was unique, we performed two experiments: one where we removed the 1.0 μm image from the training data (Fig. 5(a)) and another where we added the 0.75 μm image to the training data (Fig. 5(b)).

In the first experiment, where the 1.0 μm image was excluded, we found that the estimation for 1.0 μm was unsuccessful, similar to the previous result with the 0.75 μm image. However, other images with sizes of 0.5, 1.5, and 2.0 μm that were included in the training data were estimated successfully. In the second experiment, where the 0.75 μm image was added to the training data, the estimation for 0.75 μm, which was previously unsuccessful, was now successful. Additionally, other image sizes included in the training data were also estimated successfully. These results indicate that the 0.75 μm image size is not inherently problematic; successful estimation can occur if the same image size is present in the training data.

To further investigate why estimation from image sizes not included in the training data, which worked well in DMI estimation, does not work for saturation magnetization, we added images with sizes of 0.6 and 0.8 μm—closer to 0.75 μm—to the training data and examined whether saturation magnetization could be estimated from the 0.75 μm image (Fig. 5(c)). As a result, compared to Fig. 4(c), we found that the estimation was successful except in regions with low saturation magnetization. This indicates that, even when estimating saturation magnetization, it is not necessary to include images of the exact same size in the training data.

When comparing the estimation of DMI and saturation magnetization, it is evident that a broader range of image sizes could successfully estimate DMI, while saturation magnetization required closer image sizes (0.6 or 0.8 μm) for successful estimation from a 0.75 μm image. This reveals a significant difference in the tolerance for image size variations depending on the type of magnetic parameter being estimated. Although the exact reason for this difference has not yet been elucidated, it appears to be closely related to the underlying mechanisms of parameter estimation and offers an important clue for further investigation. Additionally, the fact that estimation from varying image sizes is possible reinforces the potential of this method as a novel approach for determining magnetic parameters.

In conclusion, we successfully estimated the DMI constant and saturation magnetization from magnetic domain images of different sizes for the first time. Our findings suggest that the AI estimates DMI based on consistent domain wall shapes that remain unchanged despite resizing, rather than relying solely on curvature or specific scale-dependent features. Additionally, we discovered a significant difference in the tolerance for image size variations depending on the type of magnetic parameter being estimated (DMI constant and saturation magnetization).

## Acknowledgements

We would like to thank Prof. Ukita, Prof. Hayashi, Dr. Yamada, Dr. Kawaguchi and Ms. Kuno for helpful discussion. This work was partially supported by the Toyoaki scholarship foundation.

## AUTHOR DECLARATIONS

### Conflict of Interest

The authors have no conflicts to disclose.

### Author Contributions

**Akito Watanabe**: Conceptualization (equal); Data Curation (lead); Formal Analysis (lead); Investigation (lead); Software (equal); Validation (lead); Writing/Original Draft Preparation (supporting), **Yoshinobu Nakatani**: Data Curation (lead); Investigation (supporting); Resources (equal); Software (equal); Writing/Review & Editing (supporting), **Hiroyuki Awano**: Investigation (supporting); Writing/Review & Editing (supporting), **Kenji Tanabe**: Conceptualization (equal); Formal Analysis (supporting); Funding Acquisition (lead); Investigation (supporting); Resources (equal); Supervision (lead); Validation(supporting); Writing/Original Draft Preparation (lead); Writing/Review & Editing (lead)

## Data Availability

The data that support the findings of this study are available from the corresponding author upon reasonable request.

## References

1.      L. Passell, O. W. Dietrich, and J. Als-Nielsen, Phys. Rev. B **14**, 4897 (1976).

2.      E. Girt, W. Huttema, O. N. Mryasov, E. Montoya, B. Kardasz, C. Eyrich, B. Heinrich, A. Yu. Dobin, and O. Karis, J. Appl. Phys. **109**, 07B765 (2011).

3.      X. Liu, M. M. Steiner, R. Sooryakumar, G. A. Prinz, R. F. C. Farrow, and G.

Harp, Phys. Rev. B **53**, 12166 (1996).

4. M. Agrawal, V. I. Vasyuchka, A. A. Serga, A. D. Karenowska, G. A. Melkov, and B. Hillebrands, Phys. Rev. Lett. **111**, 107204 (2013).

5. C. J. Safranski, Y.-J. Chen, I. N. Krivorotov, and J. Z. Sun, Appl. Phys. Lett. **109**, 132408 (2016).

6. G. A. Riley, J. M. Shaw, T. J. Silva, H. T. Nembach, Appl. Phys. Lett. **120**, 112405 (2022).

7. I. E. Dzyaloshinskii, Sov. Phys. JETP **5**, 1259 (1957).

8. T. Moriya, Phys. Rev. **120**, 91 (1960).

9. K.-S. Ryu, L. Thomas, S.-H. Yang, and S. Parkin, Nat. Nanotech. **8**, 527 (2013).

10. S. Emori, U. Bauer, S.-M. Ahn, E. Martinez, and G. S. D. Beach, Nature Mater. **12**, 611 (2013).

11. S. Pizzini, J. Vogel, S. Rohart, L. D. Buda-Prejbeanu, E. Jué, O. Boulle, I. M. Miron, C. K. Safeer, S. Auffret, G. Gaudin, and A. Thiaville, Phys. Rev. Lett. **113**, 047203 (2014).

12. S. Kim, P.-H. Jang, D.-H. Kim, M. Ishibashi, T. Taniguchi, T. Moriyama, K.-J. Kim, K.-J. Lee, and T. Ono, Phys. Rev. B **95**, 220402(R) (2017).

13. K. Di, V. Li Zhang, H. S. Lim, S. C. Ng, M. H. Kuok, J. Yu, J. Yoon, X. Qiu, and H. Yang, Phys. Rev. Lett. **114**, 047201 (2015).

14. J. Cho, N.-H. Kim, S. Lee, J.-S. Kim, R. Lavrijsen, A. Solignac, Y. Yin, D.-S. Han, N. J. J. van Hoof, H. J. M. Swagten, B. Koopmans and C.-Y. You, Nature Communications **6**, 7635 (2015)

15. A. K. Chaurasiya, C. Banerjee, S. Pan, S. Sahoo, S. Choudhury, J. Sinha and A. Barman, Scientific Reports **6**, 32592 (2016).

16. M. Kuepferling, A. Casiraghi, G. Soares, G. Durin, F. Garcia-Sanchez, L. Chen, C. H. Back, C. H. Marrows, S. Tacchi, and G. Carlotti, Rev. Mod. Phys. **95**, 015003 (2023).

17. A. Thiaville, S. Rohart, E. Jue, V. Cros, and A. Fert, Europhys. Lett. **100**, 57002 (2012).

18. A. Fert, V. Cros, and J. Sampaio, Nature Nanotech. **8**, 152 (2013).

19. L. Caretta, M. Mann, F. Büttner, K. Ueda, B. Pfau, C. M. Günther, P. Hessing, A. Churikova, C. Klose, M. Schneider, D. Engel, C. Marcus, D. Bono, K. Bagschik,

S. Eisebitt, and G. S. D. Beach, Nature Nanotechnol. **13**, 1154 (2018).

20. K. Cai, Z. Zhu, J. M. Lee, R. Mishra, L. Ren, S. D. Pollard, P. He, G. Liang, K. L. Teo, and H. Yang, Nature Electron. **3**, 37 (2020).

21. K. Tanabe and K. Yamada, Applied Physics Express **11**, 113003 (2018).

22. S. Ranjbar, S. Sumi, K. Tanabe, and H. Awano, Materials Advances **3**, 7028 (2022).

23. K. Tanabe: *Future method for estimating parameters in magnetic films* (the 68th Annual Conference on Magnetism and Magnetic Materials, 2023).

24. D. Wang, S. Wei, A. Yuan, F. Tian, K. Cao, Q. Zhao, Y. Zhang, C. Zhou, X. Song, D. Xue, and S. Yang, Advanced Science **7**, 2000566 (2020).

25. H. Y. Kwon, H. G. Yoon, C. Lee, G. Chen, K. Liu, A. K. Schmid, Y. Z. Wu, J. W. Choi, and C. Won, Science Advances **6**, eabb0872 (2020).

26. M. Kawaguchi, K. Tanabe, K. Yamada, T. Sawa, S. Hasegawa, M. Hayashi, and N. Nakatani, npj Computational Materials **7**, 20 (2021).

27. H. Y. Kwon, H. G. Yoon, S. M. Park, D. B. Lee, J. W. Choi, and C. Won, Advanced Science **8**, 2004795 (2021).

28. D. B. Lee, H. G. Yoon, S. M. Park, J. W. Choi, H. Y. Kwon, and C. Won, Scientific Reports **11**, 22937 (2021).

29. N. Mehmood, J. Wang, and Q. Liu, J. Appl. Phys. **132**, 043904 (2022).

30. S. Kuno, S. Deguchi, S. Sumi, H. Awano, and K. Tanabe, APL Machine Learning **1**, 046111 (2023).

31. A. Hubert and R. Schaefer, *Magnetic Domains: The Analysis of Magnetic Microstructures* (Springer, 1998).

32. Y. Nakatani, A. Thiaville, and J. Miltat, Nature Mater. **2**, 521 (2003).

33. T. Sato and Y. Nakatani, J. Magn. Soc. Jpn. **35**, 163 (2011).

**Caption**

**Figure 1**

Fig. 1. (a) Schematic image of a method for estimating magnetic parameters by using machine learning. There is a software including the trained AI. The AI learned a huge number of datasets of a magnetic domain image and magnetic parameters, which are generally generated by a simulation. The software can estimate the parameters from other magnetic domain images. A researcher fabricates a magnetic film and takes a magnetic domain image of the film by a microscope. When the image input to the software, he/she can obtain magnetic parameters as an output from the software. (b) Schematic images of making images with different imaging size. The original images are made by a micromagnetic simulation. The size is $2.0 \times 2.0 \ \mu m^2$ and the pixel size is $512 \times 512$. The images are changed from the image size of $2 \times 2 \ \mu m^2$ into other image sizes such as $0.5 \times 0.5$, $1.0 \times 1.0$, and $1.5 \times 1.5 \ \mu m^2$ by cutting. After that, the pixel size in all the images is changed into $128 \times 128$.

**Figure 2**

Fig. 2. (a) Experimental conditions of train and test data. The train data are data of one size and three trained AIs are prepared using the data of $0.5 \times 0.5$, $1.0 \times 1.0$, and $1.5 \times 1.5 \ \mu m^2$. All the test data are the image with $2.0 \times 2.0 \ \mu m^2$ size. (b) Estimation results of $D$ values. Cross, triangle and circle denote using train data of $0.5 \times 0.5$, $1.0 \times 1.0$, and $1.5 \times 1.5 \ \mu m^2$ sizes, respectively. (c-d) Estimation results as image sizes of the test data are equal to those of the train data. The image sizes are $0.5 \times 0.5$ (c) and $1.0 \times 1.0 \ \mu m^2$(d).

**Figure 3**

Fig. 3. (a) Experimental conditions of train and test data. The train data are data of $0.5$, $1.0$, $1.5$, and $2.0 \ \mu m$. The test data are data of $0.5$, $1.0$, $1.5$, $2.0$, $0.75$, $1.25$, and $1.75 \ \mu m$. These data are separated into Tests A($0.5$, $1.0$, $1.5$, and $2.0 \ \mu m$) and B($0.75$, $1.25$, and $1.75 \ \mu m$). (b-c) Estimation results of the DMI constants in Tests A(b) and B(c).

**Figure 4**

Fig. 4. (a) Experimental conditions of train and test data. The train data are data of $0.5$, $1.0$, $1.5$, and $2.0 \ \mu m$. The test data are data of $0.5$, $1.0$, $1.5$, $2.0$, $0.75$, $1.25$, and $1.75 \ \mu m$. These data are separated into Tests A($0.5$, $1.0$, $1.5$, and $2.0 \ \mu m$) and B($0.75$, $1.25$, and $1.75 \ \mu m$) like Fig. 3(a). (b-c) Estimation results of the saturation magnetization in Tests A(b) and B(c). (d) Relation between the saturation magnetization and magnetic domain size. Larger saturation magnetization generates smaller magnetic domain size.

**Figure 5**

Fig. 5. (a-c) Experimental conditions of train and test data, and the estimation results of the saturation magnetization. The train data are the data of $0.5$, $1.5$, and $2.0$ μm and the test data are the data of $0.5$, $1.0$, $1.5$, and $2.0$ μm(a). The train data are the data of $0.5$, $0.75$, $1.5$, and $2.0$ μm and the test data are the data of $0.5$, $0.75$, $1.5$, and $2.0$ μm(b). The train data are the data of $0.5$, $0.6$, $0.8$, and $1.0$ μm and the test data are the data of $0.75$ μm(c).

**Table 1**

Table 1. Summary of the magnetic parameter values used in the simulation. $\sigma$ is magnetic anisotropic dispersion, which is defined by $\Delta K_u / K_u$. Here $\Delta K_u$ is its variation, which represents pinning sites due to roughness, defects and impurities in a ferromagnetic layer.

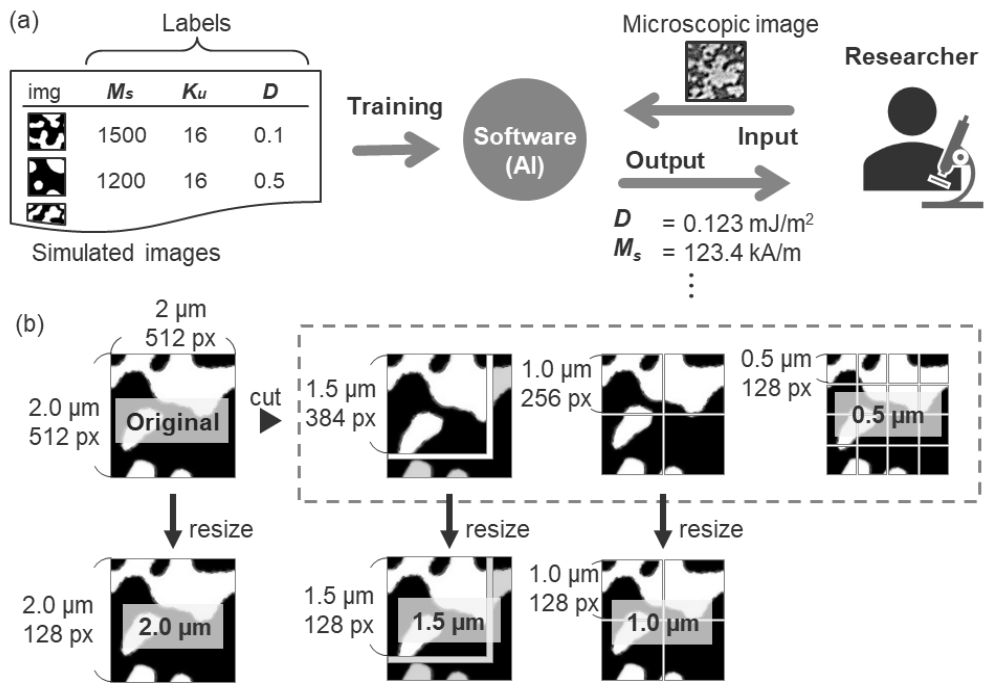

Fig. 1

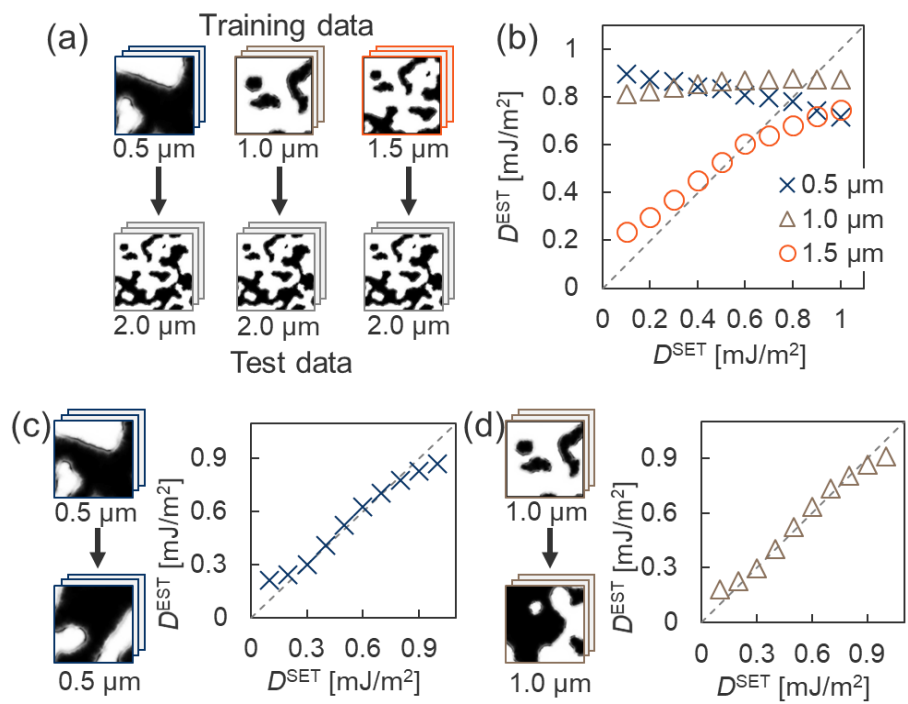

Fig. 2

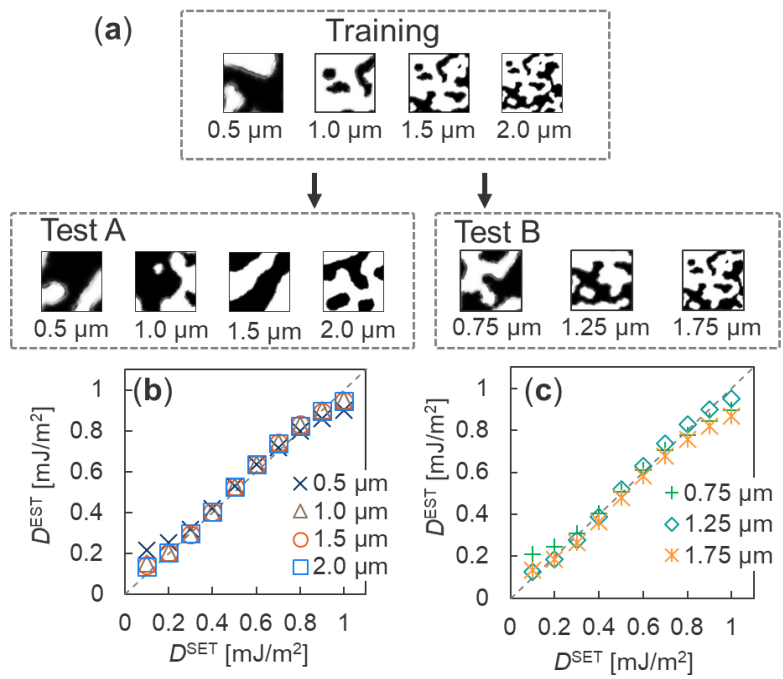

Fig. 3

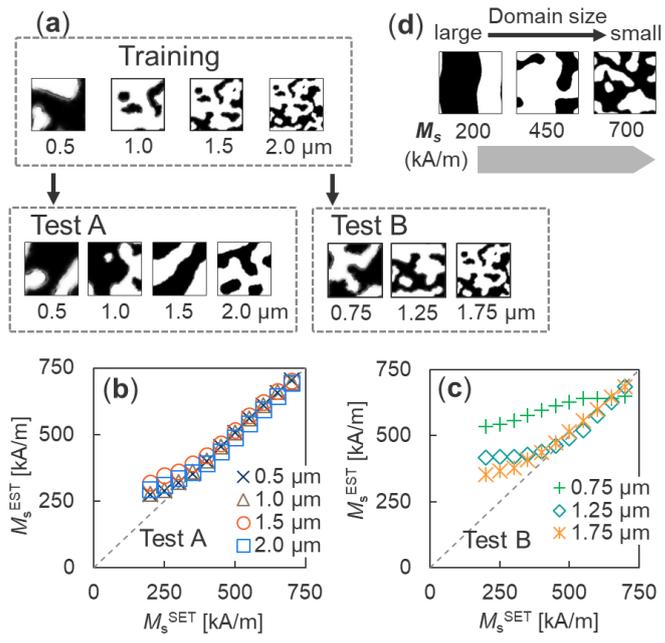

Fig. 4

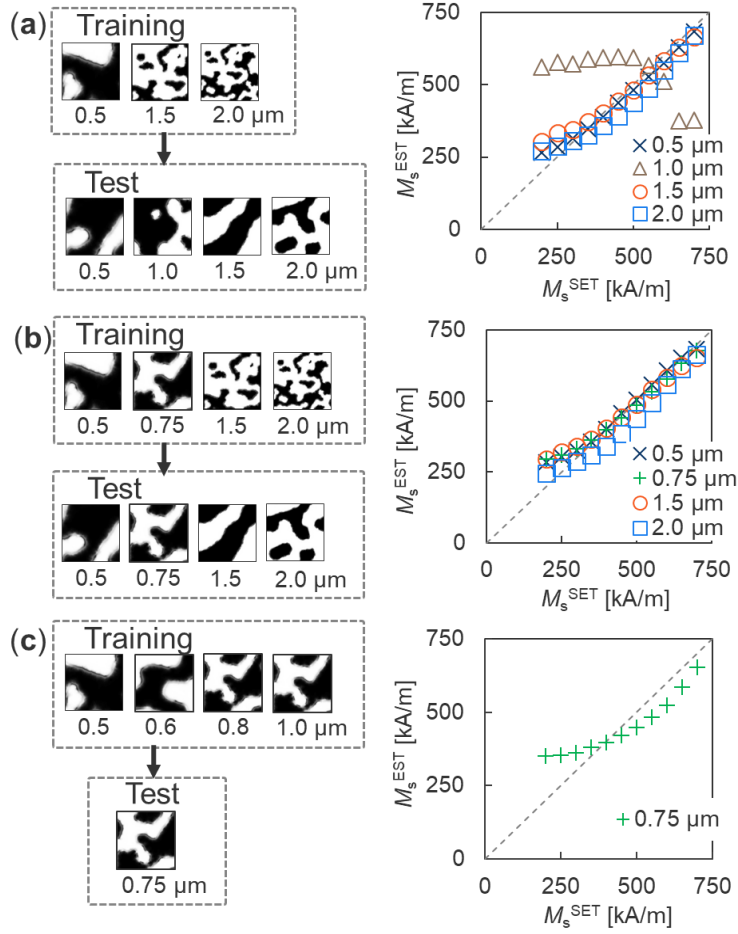

Fig. 5

|          | $M_s$ [kA/m] | $K_u$ [MJ/m$^3$] | $A_{ex}$ [pJ/m] | $D$ [mJ/m$^2$] | $\sigma$ |
|----------|-------------|-----------------|-----------------|----------------|----------|
| Figs. 2-3 | 1,500 | 1.6 | 31 | $0.1 - 1.0$ | $0.05 - 0.2$ |
| Figs. 4-5 | $200 - 700$ | 0.4 | 10 | 0 | $0.05 - 0.2$ |

Table 1